\documentclass{jfm}

\usepackage{graphicx}
\usepackage{natbib}
\usepackage{amsmath}
\usepackage{hyperref}

\usepackage[usenames,dvipsnames]{color}
	\def\jya#1{\textcolor{black}{#1}}		
	\def\jyb#1{\textcolor{black}{#1}}		
\ifCUPmtlplainloaded \else
  \checkfont{eurm10}
  \iffontfound
    \IfFileExists{upmath.sty}
      {\typeout{^^JFound AMS Euler Roman fonts on the system,
                   using the 'upmath' package.^^J}%
       \usepackage{upmath}}
      {\typeout{^^JFound AMS Euler Roman fonts on the system, but you
                   dont seem to have the}%
       \typeout{'upmath' package installed. JFM.cls can take advantage
                 of these fonts,^^Jif you use 'upmath' package.^^J}%
      }
  \else
  \fi
\fi

\ifCUPmtlplainloaded \else
  \checkfont{msam10}
  \iffontfound
    \IfFileExists{amssymb.sty}
      {\typeout{^^JFound AMS Symbol fonts on the system, using the
                'amssymb' package.^^J}%
       \usepackage{amssymb}%
       \let\le=\leqslant  
       \let\ge=\geqslant  
      }{}
  \fi
\fi

\ifCUPmtlplainloaded \else
  \IfFileExists{amsbsy.sty}
    {\typeout{^^JFound the 'amsbsy' package on the system, using it.^^J}%
     \usepackage{amsbsy}}
    {}
\fi

\newsavebox{\astrutbox}
\sbox{\astrutbox}{\rule[-5pt]{0pt}{20pt}}

\title[Statistical behaviour of self-similar structures in canonical wall turbulence]{Statistical behaviour of self-similar structures in canonical wall turbulence}

\author[J. Hwang, J. H. Lee and H. J. Sung]%
{Jinyul Hwang$^1$\thanks{Email address for correspondence: jhwang@pusan.ac.kr}
, Jae Hwa Lee$^2$\thanks{Email address for correspondence: jhlee06@unist.ac.kr}
and Hyung Jin Sung$^3$}

\affiliation{$^1$School of Mechanical Engineering, Pusan National University, 
2 Busandaehak-ro 63beon-gil, Geumjeong-gu, Busan 46241, Korea\\[\affilskip]
$^2$Department of Mechanical Engineering, UNIST, 
50 UNIST-gil, Ulsan 44919, Korea\\[\affilskip]
$^3$Department of Mechanical Engineering, KAIST, 
291 Daehak-ro, Yuseong-gu, Daejeon 34141, Korea\\[\affilskip]
}

\pubyear{2010}
\volume{650}
\pagerange{119--126}
\date{4 April 2020; revised 11 July 2020; accepted 18 August 2020.}
\begin{document}

\maketitle

\begin{abstract}
Townsend's attached-eddy hypothesis (AEH) provides a theoretical description of turbulence statistics in the logarithmic region in terms of coherent motions that are self-similar with the wall-normal distance ($y$).
This hypothesis was further extended by Perry and coworkers who proposed attached-eddy models that predict the coexistence of the logarithmic law in the mean velocity and streamwise turbulence intensity as well as spectral scaling for the streamwise energy spectra.
The AEH can be used to predict the statistical behaviours of wall turbulence, yet revealing such behaviours has remained an elusive task because the proposed description is established within the limits of asymptotically high Reynolds numbers.
\jya{Here, we show the self-similar behaviour of turbulence motions contained within wall-attached structures of streamwise velocity fluctuations using the direct numerical simulation dataset of turbulent boundary layer, channel, and pipe flows ($Re_\tau \approx 1000$).}
The physical sizes of the identified structures are geometrically self-similar in terms of height, and the associated turbulence intensity follows the logarithmic variation in all three flows.
Moreover, the corresponding two-dimensional energy spectra are aligned along a linear relationship between the streamwise and spanwise wavelengths ($\lambda_x$ and $\lambda_z$, respectively) in the large-scale range ($12y < \lambda_x <$ 3--4$\delta$), which is reminiscent of self-similarity.
Consequently, \jya{one-dimensional spectra obtained by integrating the two-dimensional spectra over the self-similar range show some evidence for self-similar scaling $\lambda_x \sim \lambda_z$ and the possible existence of $k_x^{-1}$ and $k_z^{-1}$ scaling regions in a similar subrange}.
The present results reveal that the asymptotic behaviours can be obtained by identifying the self-similar coherent structures in canonical wall turbulence, albeit in low Reynolds number flows.
\end{abstract}


\section{Introduction}\label{sec:intro}
One of the distinct features of wall-bounded turbulent flows is that they are characterised by multiple scales over a broad range.
Owing to the presence of a solid wall, the characteristic length scales in wall turbulence vary from the viscous length scale ($\delta_\nu$) to the outer length scale ($\delta$).
Such a multiscale nature is described by the friction Reynolds number ($Re_\tau$), which is the ratio of $\delta$ and $\delta_\nu$.
At asymptotically high $Re_\tau$, there is a region where two different length scales are valid simultaneously.
In this region, the only relevant length scale is the distance from the wall $y$, and the mean velocity profile follows the logarithmic variation with respect to $y$ \citep{Millikan38}; this is the so-called logarithmic region.
\cite{Townsend76} conjectured that energy-containing motions in the logarithmic region are self-similar and that their sizes are proportional to $y$ because the impermeability of the walls restricts the size of the order of the wall-normal distance.
In this respect, Townsend described that these motions are attached to the wall, which is the so-called attached-eddy hypothesis (AEH); see a recent review \citep{Marusic19}.
The AEH allows us to establish the relationship between coherent structures and turbulence statistics at asymptotically high $Re_\tau$ by conjecturing that the logarithmic region is composed of self-similar energy-containing motions.
These motions are assumed to be inviscid near the wall; they lead to the logarithmic variation in the wall-parallel components of the turbulence intensities and constant wall-normal turbulence intensity over the logarithmic region.

The AEH was further extended by \cite{Perry82} who deduced the attached-eddy model. 
They showed that the eddies are randomly distributed in a hierarchical form with a probability density function (PDF) that is inversely proportional to $y$ because of their self-similar nature.
Based on the attached-eddy model, they expected that the logarithmic variations occur both in the mean velocity and in the wall-parallel components of turbulence intensity.
Moreover, the model also shows that self-similar motions contribute to a $k_x^{-1}$ scaling in the one-dimensional spectra of the streamwise velocity, where $k_x$ is the streamwise wavenumber.
The $k_x^{-1}$ region was predicted by \cite{Perry77}, who assumed that there is a spectral overlap region where $y$ and $\delta$ scalings hold simultaneously.
In this sense, the $k_x^{-1}$ region can be a consequence of the AEH and is deemed the spectral signature of attached eddies \citep{Perry82,Perry86}.
However, although some studies have reported empirical evidence for the existence of the $k_x^{-1}$ region \citep{Nickels05,Vallikivi15}, it remains unclear whether such scaling exists in a high Reynolds number flow \citep{Rosenberg13,MK15,Ahn15,Agostini17,Chandran17,Baars20}.
This leads to questions regarding the relationship between the $k_x^{-1}$ region and the logarithmic variation of the streamwise turbulence intensity because the latter is satisfied when the spectral overlap argument is valid.
Moreover, several high-Reynolds-number experiments have revealed the logarithmic variations in the streamwise turbulence intensity \citep{Hultmark12,Marusic13,Orlu17}. 
However, the $k_x^{-1}$ dependence in the same experimental data has not been yet established, especially in premultiplied one-dimensional spectra with a semi-log plot \citep{Baars20}.

\jya{The ambiguities in the spectral signatures of the attached eddies} are related to insufficient scale separation since the AEH requires that the Reynolds number approaches infinity.
This means that there is no region where both $y$ and $\delta$ scalings are valid over the same wavenumber space, even in the high Reynolds number experiments $Re=O(10^{4-5})$ of the aforementioned studies.
However, given the fact that the self-similar energy-containing motions follow a hierarchical distribution \citep{Perry82}, we may expect that self-similar motions can exist even if scale separation is insufficient. 
Owing to insufficient separation of scales, self-similar motions are less statistically dominant than other coexisting motions, which in turn leads to ambiguity of the asymptotic behaviours in turbulent statistics.
If we properly extract the contributions of self-similar motions by filtering out coexisting motions in turbulent flows, then we may observe the spectral overlap region \jya{\citep{Baars20}}.
This conjecture is supported by the existence of very-large-scale structures \citep{Kim99} or global modes \citep{Del04} in internal flows and superstructures in external flows \citep{Hutchins07}.
The very large scales, characterised by the outer scale $\delta$, extend from the outer region to the near-wall region and significantly contribute to turbulence statistics in the logarithmic region \citep{Guala06,Balakumar07,Lee11,Hwang16}.
In other words, the statistical behaviours of self-similar motions in the logarithmic region can be contaminated by the contributions of non-self-similar motions.
\cite{Jimenez08} showed that the departure of the streamwise turbulent intensity from the logarithmic variation is due to the contamination by very long and wide motions (i.e. global modes).
It is to be noted that other types of coexisting motions related to smaller scales (e.g. viscous scales) also contribute to the absence of the logarithmic law \citep{Perry86,Perry90,Marusic97}, but their effect on the ambiguity of the $k_x^{-1}$ scaling may be negligible because such a scaling region will appear over a large scale range $O(y)$.

Several works have shown statistical evidence for the existence of self-similar motions in the logarithmic region.
\cite{Del06b} and \cite{Lozano12} extracted  three-dimensional clusters of vortices and sweeps/ejections in instantaneous flow fields of channel flows and showed that the sizes of the wall-attached objects are proportional to the height.
In an artificial channel flow, which resolves only a given spanwise length scale, it was found that the statistical behaviours of energy-containing motions in the logarithmic region are characterised by the spanwise length scale and are self-similar with respect to $y$ \citep{Hwangy15}.
Similarly, \cite{Hellstrom16} reported that the azimuthal length scales of the dominant modes, identified by a proper orthogonal decomposition analysis, are linearly proportional to $y$ over a decade in turbulent pipe flows.
In turbulent boundary layers (TBL), \cite{Baars17} extracted coherent motions using a spectral coherence analysis and showed that the corresponding motions follow a constant streamwise/wall-normal ratio.
Subsequently, a linear relationship between the streamwise and spanwise wavelengths, which is evidence for self-similarity, was observed in the two-dimensional spectra of streamwise velocity \citep{Chandran17}.
\cite{Agostini17} found that the premultiplied derivative of the second-order structure function has a constant region, which reflects the $k_x^{-1}$ scaling \citep{Davidson06}.
However, although previous studies have shown the existence of self-similar coherent motions in wall turbulence, the relationship between the identified motions and statistical behaviours (i.e. logarithmic variation or $k_x^{-1}$ region) predicted by the AEH has not been established.
This is because the AEH originated from an attempt to explain the asymptotic behaviours of turbulence statistics, and especially two-point correlations, in terms of coherent structures in the logarithmic region.
Hence, there remains a need to show whether self-similar coherent motions extracted by a particular method can be attributed to the logarithmic variation in the turbulence intensity or the spectral signature for the $k_x^{-1}$ scaling and to reveal whether such behaviours appear in a consistent range where the identified motions are defined.

\jya{Recently, \cite{Srinath18} proposed a model of the one-dimensional streamwise energy spectrum based on wall-attached structures of streamwise velocity fluctuations ($u$) identified in a streamwise--wall-normal plane.
They showed that both streamwise energy spectrum and turbulence intensity follow a power-law distribution with respect to the streamwise length scale, and the sum of the corresponding power-law exponents is close to $-1$.
In particular, the exponent of the streamwise energy spectrum becomes $-1$ (i.e. $k_x^{-1}$ scaling) over $100 < y^+ < 200$ (below the logarithmic region) at $Re_\tau \approx O(10^{3-4})$.
By extracting a spine or skeleton of contiguous volumes of the intense $u$ regions, \cite{Solak18} examined the geometrical features of $u$ structures at $Re_\tau \approx 700$.
It was found that the conditionally sampled streamwise intensity follows a power-law distribution with respect to the structure length at a given $y$.
In addition, the sum of the exponents of the one-dimensional spectra and the intensity was observed close to $-1$, which supports the model proposed by \cite{Srinath18}.}

\cite{Hwang18} demonstrated that the wall-attached structures of $u$ are not only self-similar in terms of their heights, ($l_y$) but also directly contribute to the logarithmic variation in the streamwise turbulence intensity.
In addition, they showed that the population density of the identified structures is inversely proportional to $l_y$, reminiscent of the hierarchies of self-similar eddies \citep{Townsend76,Perry82}.
It is worth mentioning that the presence of the logarithmic region in the reconstructed intensity profile was verified by an apparent plateau in the indicator function of the logarithmic variation although there is no logarithmic behaviour in the profile of the streamwise turbulence intensity at a low Reynolds number ($Re_\tau \approx 1000$).
Moreover, in pipe flows, it was found that the profile of the streamwise velocity reconstructed by the superposition of the wall-attached $u$ structures exhibits the logarithmic variation \citep{Hwang19}.
In particular, the ranges of the logarithmic variation in the streamwise turbulence intensity and in the mean velocity appear at a consistent location $3Re_\tau^{1/2} < y^+ < 0.18\delta^+$, where the superscript $+$ denotes viscous scaling.
These findings support the supposition that the identified $u$ structures can be regarded as the structural basis of the logarithmic region in the context of the AEH.

Despite evidence on the self-similarity of wall-attached $u$ structures and their contribution to the logarithmic behaviours, their spectral contribution and the turbulence motions contained within the identified structures have not been revealed.
The wall-attached $u$ structures are identified in instantaneous fluctuating velocity fields and their length scales are measured in terms of the dimensions of the bounding box of each object in physical space.
The length and width of the wall-attached structure does not necessarily indicate a particular length scale associated with the identified object since each structure can contribute turbulence energy over a certain range of spectral space \citep{Nickels01}. 
In other words, the physical size of each structure is a particular length scale among a range of scales contained within the individual structure; in particular, it represents one of the large scales related to the volume of intense $u$.
Hence, spectral analysis is required to reveal whether the large scales contained within the identified object can be attributed to the logarithmic variation or exhibit the spectral overlap argument proposed by Perry and coworkers \citep{Perry77,Perry82,Perry86}.

Given the self-similar nature of the wall-attached $u$ structures, we may expect to see a linear relationship between the streamwise and spanwise wavelength ($\lambda_x$ and $\lambda_z$) in the two-dimensional energy spectra of $u$ \citep{Chung15,Chandran17,Deshpande20}.
In order for there to be a $k_x^{-1}$ region in the one-dimensional spectra, the two-dimensional energy spectra should be characterised by a region of constant energy in the logarithmic region, and this region should be bounded by an identical power-law between $\lambda_x$ and $\lambda_z$ \citep{Chung15}.
\cite{Chandran17} observed that constant-energy contours in the two-dimensional spectra are bounded by a linear relationship ($\lambda_x \sim \lambda_z$), which represents the self-similarity of turbulence motions.
They found that the linear behaviour appears at high $Re_\tau (\approx 26000)$ TBLs in the large-scale range ($\lambda_x > 10y$), while only square root behaviour is observed at low $Re_\tau (\approx 2400)$; this is consistent with the work of \cite{Del04} who examined the two-dimensional energy spectra of a turbulent channel flow at $Re_\tau = 1900$.
However, there was no clear plateau in the premultiplied one-dimensional spectra (i.e. $k_x^{-1}$ region) over a similar range of the linear behaviour in the two-dimensional spectra.
As discussed, the energy spectra of $u$ represent the contributions of all coexisting eddies and this obscures the appearance of self-similar behaviours since those behaviours are only achieved in the limit of infinite $Re_\tau$.
Therefore, extracting proper $u$ motions satisfying the AEH is required to examine the spectral-overlap arguments.
Recent studies have been conducted to filter out the spectral contribution of self-similar motions.
Given the hierarchies of the attached eddies, \jya{\cite{Hu20} extracted $u$ in the one-dimensional spectra over the range of $5.7y < \lambda_x < 3-4\delta$ and $y^+ > 100$.}
\cite{Baars20} extracted the energy contributed from wall-coherent motions by decomposing the streamwise turbulence intensity through spectral coherence analysis. 
Using the two-point correlation of $u$, \cite{Deshpande20} obtained the two-dimensional spectra, which represent the energy distribution of the wall-attached motions across $\lambda_x$ and $\lambda_z$.
Although these studies showed that the extracted energy distributions exhibit the self-similarity ($\lambda_x \sim y$ or $\lambda_x \sim \lambda_z$) \jya{in the logarithmic region}, the extracted signals also include contributions from tall wall-attached motions (non-self-similar motions) that extend from the wall to the outer region \jya{(i.e. beyond the logarithmic region)}.
In other words, contamination by non-self-similar motions could mask the asymptotic statistical behaviours predicted by the AEH.

The objective of the present study is to explore the spectral contribution of turbulence motions that comprise wall-attached $u$ clusters by computing the two-dimensional spectra of $u$, in which the velocity signals contained within self-similar structures are isolated.
To do so, we examine the direct numerical simulation (DNS) data of a fully developed turbulent channel and pipe flows, along with zero-pressure-gradient TBL at $Re_\tau \approx 1000$, and identify the wall-attached self-similar structures by applying universal filters in terms of height.
The wall-attached self-similar clusters, identified in the physical space, not only exhibit similar geometrical features but also embody scales corresponding to $\lambda_x \sim \lambda_z$, which in turn contributes to the existence of the  $k_x^{-1}$ and $k_z^{-1}$ scalings.
A brief description of the DNS data and the identification method for extracting self-similar clusters in instantaneous flow fields is provided in \S \ref{sec:method}.
In \S \ref{sec:sec1}, wall-attached $u$ clusters are decomposed into the buffer-layer, self-similar, and non-self-similar structures in terms of their height.
Next, the wall-attached self-similar structures are examined using the two-dimensional energy spectra to reveal the self-similar behaviours in the logarithmic region of all three flows.
We then explore the one-dimensional streamwise and spanwise spectra by comparing the energy contained in the spectral range where the energetic ridges in the two-dimensional spectra follow a linear relationship between the wall-parallel wavelengths.
Finally, a summary of the main findings is provided in \S \ref{sec:Concl}.

\section{DNS data and cluster identification method}\label{sec:method}
In this study, the DNS data of the zero-pressure gradient TBL \citep{Hwang17,Yoon18}, and the fully developed turbulent channel \citep{Lee14,Lee15} and pipe flows \citep{Ahn13,Ahn15} are analysed.
To solve the Navier--Stokes equations for incompressible flow, the DNS was performed using the fractional step method proposed by \citet{Kim02}.
Table \ref{tab:tb1} provides the parameters of the DNS data, and a detailed description of the DNS can be found in the aforementioned studies.
The friction Reynolds numbers, defined as the ratio of the outer length scale to the viscous length scale, are matched at $Re_\tau = u_{\tau}\delta/\nu \approx 1000$.
Here, $u_\tau$ is the friction velocity, $\nu$ is the kinematic velocity, and $\delta$ is the flow thickness (i.e. channel half-height, pipe radius or $99\%$ boundary layer thickness).
Throughout the present work, the superscript $+$ represents viscous scaling ($\nu/u_\tau$ and $u_\tau$).
The friction Reynolds number of the TBL ($Re_\tau = 980$) is chosen at the middle of the streamwise length ($L_x \approx 11.7\delta$) of the subdomain.
We neglect the Reynolds-number effect since $Re_\tau$ varies from 913 to 1039 across the streamwise direction of the extracted flow field.
In the present study, $x$, $y$, and $z$ indicate the streamwise, wall-normal, and spanwise directions, respectively.
In the pipe flow, the wall-normal direction is defined as $y = \delta - r$, where $r$ denotes the radial direction.
In addition, for an analogy with $z$ in the TBL and channel flows, the spanwise dimension of the pipe is defined as the arclength $r\theta$, where $\theta$ denotes the azimuthal direction.
We define the streamwise velocity fluctuations $u = U - \overline{U}(y)$, where $U$ is the streamwise velocity, and the overbar denotes the ensemble average.
For the TBL, the streamwise fluctuating component is decomposed by considering the local height of the turbulent/non-turbulent interface $\delta_i$ \citep{Kwon16}; that is, $\tilde{u} = U - \tilde{U}(y,\delta_i)$, where $\tilde{U}$ is the conditional mean velocity as a function of $y$ and $\delta_i$.
The profile of $\tilde{U}$ shows a significant discrepancy compared with that of $\overline{U}$ in the intermittent region, whereas they collapse close to the wall \citep{Kwon16,Yoon20}.
We focus on coherent motions in the logarithmic region, and thus the fluctuating fields are insensitive to the decomposition method (i.e. $u \approx \tilde{u}$ in TBL).
In other words, the results presented here remain qualitatively unchanged when using the Reynolds decomposition.
Hereafter, we refer to $\tilde{u}$ as $u$ in the TBL.
However, when we examine fluctuating motions that reach $\delta_i$ or reside in the intermittent region, we have to consider the oscillation of $\delta_i$ in order to avoid contamination in the intermittent region of TBL.
\begin{table}
  \begin{center}
\def~{\hphantom{0}}
  \begin{tabular}{cccccccccc}
      Case & $Re_{\tau}$ & $(L_x, L_y, L_z)$ & $(N_x, N_y, N_z)$ & $\Delta x^+$ & $\Delta y^+_{min}$ & $\Delta y^+_{max}$ & $\Delta z^+$ & $\Delta t^+$ \\[3pt]
      TBL     & $980$ & $(2300\delta_0, 100\delta_0, 100\delta_0)$ & $(13313, 541, 769)$ & $5.49$ & $0.159$  & $9.56$ & $4.13$ & $0.0504$ \\[3pt]
     \textcolor{BrickRed}{Channel} & $930$ & $(10\pi\delta, 2\delta, 3\pi\delta)$               & $(4993, 401, 2497)$ & $5.86$ & $0.0287$ & $7.31$ & $3.51$ & $0.0618$ \\[3pt]
     \textcolor{Cerulean}{Pipe}      & $930$ & $(30\delta, 2\delta, 2\pi\delta)$                   & $(4097, 301, 1025)$ & $6.84$ & $0.166$   & $9.24$ & $5.73$ & $0.246$ \\[3pt]
  \end{tabular}
  \caption{Simulation parameter. Here, $Re_\tau$ is the friction Reynolds number;
$L_i$ and $N_i$ indicate the domain size and number of grid points, respectively;
grid spacings in the wall-parallel directions are $\Delta x^+$ and $\Delta z^+$;
the minimum and maximum grid sizes in the wall-normal direction are $\Delta y^+_{min}$ and $\Delta y^+_{max}$, respectively, and $\Delta t^+$ is the time step.
In the pipe dataset, the arc length is used to denote the spanwise direction.
In the turbulent boundary layer (TBL), inner-normalised resolutions are taken at $Re_{\tau} \approx 1000$ and $\delta_0$ denotes the inlet boundary layer thickness.}
  \label{tab:tb1}
  \end{center}
\end{table}

We identify the clusters of $u$ in the instantaneous flow fields by extracting the contiguous points of the intense $u$ region \citep{Hwang18,Hwang19,Han19,Yoon20}.
In the three-dimensional flow field, the irregular shapes of the objects are defined as 
\begin{equation}
u(\textbf{\textit{x}}) > \alpha u_{rms}(y) \; \text{or} \; u(\textbf{\textit{x}}) > -\alpha u_{rms}(y),
  \label{eq:eq1}
\end{equation}
where $u_{rms}$ is the standard deviation of the streamwise velocity and $\alpha$ is the threshold. 
Individual objects are extracted using the connectivity rule, in which nodes are labelled among the six orthogonal neighbours of each node satisfying (\ref{eq:eq1}) in Cartesian coordinates \citep{Moisy04,Del06b,Lozano12,Hwang18} and cylindrical coordinates \citep{Hwang19,Han19}.
We chose the threshold $\alpha = 1.5$ for all three flows; further discussion can be found in our previous works \cite{Hwang18} and \cite{Yoon20}.
In the vicinity of this value, turbulence clusters show the percolation behaviours over a wide range of $Re_\tau$ \citep{Del06b,Lozano12,Hwang19} and for different flow configurations; that is, pipes in \cite{Hwang19,Han19} or adverse-pressure-gradient TBL in \cite{Yoon20}.
The present identification method allows us to measure the physical length scales of individual structures in instantaneous flow fields.
Each structure is bounded by a box of size $l_x \times l_y \times l_z$, where the corresponding length, height, and width are denoted by $l_x$, $l_y$, and $l_z$.
In the pipe flow, $l_z$ is computed in terms of the maximum arc length in the plane obtained by projecting each object onto the cross-stream plane.

\section{\label{sec:sec1}Results and discussion}
As mentioned, we focus on wall-attached self-similar structures (WASS) of $u$ reported in the works of \cite{Hwang18} and \cite{Hwang19}.
Here, we briefly summarise the spatial characteristics and the physical interpretation of WASS.
The $u$ clusters (\ref{eq:eq1}) can be classified into wall-attached and wall-detached by measuring the minimum wall-normal distance ($y_{min}$) of each object.
Wall-attached $u$ clusters are defined as $y_{min}^+ \approx 0$, meaning that each cluster has coherence near the wall.
\jya{The attached eddies proposed by Townsend do not necessarily adhere to the wall, as the AEH considers coherent motions in asymptotically high $Re_\tau$.
Hence, these eddies are assumed inviscid, which results in non-zero wall-parallel velocity components near the wall.
Given that the AEH is an inviscid theory, we could define the wall-attached $u$ structures with $y_{min}^+ < 20$ (beyond the viscous sublayer), as in the study by \cite{Del06b} or \cite{Lozano12}.
However, we found that approximately 90\% of the wall-attached $u$ structures extending below $y^+ < 20$ have their minimum wall distance close to zero \citep{Hwang18} and, in particular, all of the identified structures that extend beyond the logarithmic region have $y_{min}^+ \approx 0$.
Because the present study primarily focuses on the contribution of the identified structures to the logarithmic region, the criterion of $y_{min}$ does not significantly affect our conclusions.}

In addition, our use of `wall-attached' and `wall-detached' is to distinguish between whether the identified clusters are physically attached to the wall or floating in a flow since each of them contains self-similar or non-self-similar structures \citep{Marusic19,Yoon20}.
This approach takes into account the description of energy-containing motions in \cite{Jimenez08}.
This nomenclature may provide a better description of coherent structures in the context of the AEH when compared with simply designating the self-similar (i.e. dimensions are proportional to $y$) structures as `attached'.
In addition, the identified wall-attached structures, i.e. those that anchor to the wall, might contribute to the skin friction since energy-containing motions in the logarithmic region are responsible for the skin friction \citep{De16,Agostini19}.

According to \cite{Hwang18}, wall-attached $u$ structures occupy approximately $90\%$ of a total volume of $u$ clusters and carry significant turbulent energy throughout the TBL.
The length and width of the wall-attached structures are scaled with their heights and the population density exhibits an inverse power-law with respect to $l_y$.
In addition, the streamwise turbulence intensity contained within these structures shows the logarithmic variation even at low $Re_\tau (\approx 1000)$, suggesting that the identified structures are prime candidates of Townsend's AEH.
Moreover, \cite{Hwang19} demonstrated that the WASS identified in turbulent pipe flows contribute to the presence of the logarithmic velocity law, and thus they may play a role as the structural basis for the inertial region.
In addition, the number of uniform momentum zones (UMZs) contained in the WASS increases with $l_y$ and the peak magnitude of the streamwise turbulence intensity in the near-wall region follows a log-linear increase with $l_y$.
These results support the inferene that the structural organisation of WASS can be interpreted in terms of both the hierarchical length-scale distribution and a nested hierarchy of the hairpin packet \citep{Adrian00}.
In this manner, the turbulence statistics carried by the WASS result from the collective contribution of coherent motions with heights of less than $l_y$; for further discussion, see \S5 in \cite{Hwang18}.

\begin{figure}
\centerline{\includegraphics[trim=0cm 0.2cm 0cm 0.2cm,width=11.5cm]{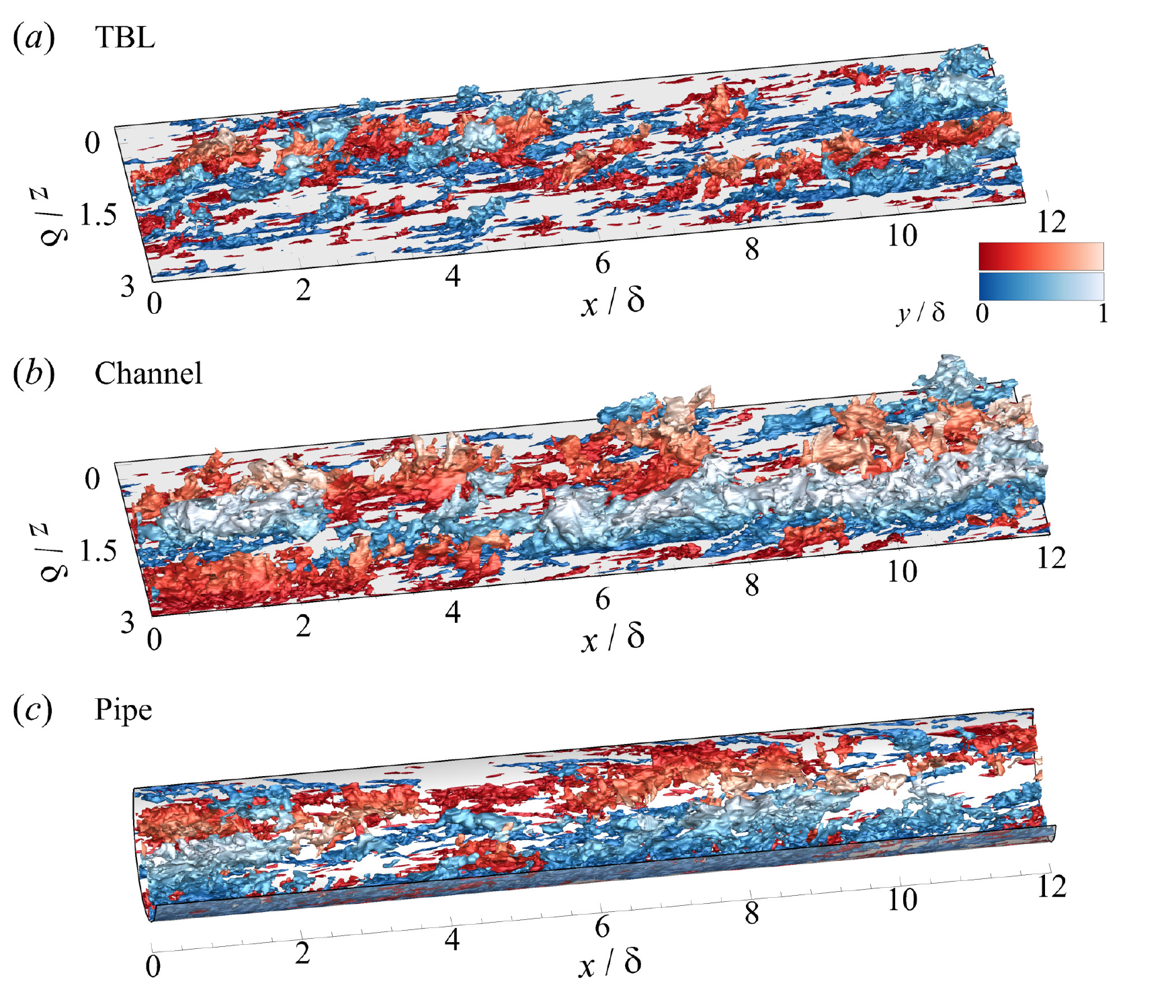}}
\caption{\label{fig:1} Isosurfaces of wall-attached structures in an instantaneous flow field: ($a$) TBL; ($b$) channel; and ($c$) pipe. Red and blue isosurfaces indicate positive and negative streamwise velocity fluctuations, respectively. Light to dark shading indicates an increase in $y/\delta$.}
\end{figure}

However, the length scales of each cluster (i.e. $l_x$ and $l_z$) correspond to the dimensions of the bounding box defined in physical space.
In other words, $l_x$ or $l_z$ is a particular length scale among a range of scales contained within the individual structure; in particular, it represents one of the large scales related to the volume of intense $u$.
We will analyse the two-dimensional energy spectrum of WASS to observe the energy distribution across the streamwise and spanwise wavelengths ($\lambda_x$ and $\lambda_z$).
This analysis allows us to clarify the wavelength scales associated with the WASS identified in the physical space and to interpret the organisation of WASS in view of the spectral argument \citep{Perry77,Perry82,Perry86}.

\subsection{Decomposition of wall-attached structures}
First, we decompose the wall-attached structures of $u$ in order to extract the self-similar structures.
Figure \ref{fig:1} shows the isosurfaces of wall-attached structures ($y_{min}^+ \approx 0$) identified in all three flows.
Here, light to dark shading indicates an increase in the wall-normal location.
As seen, we can observe the coexisting streaky structures over a wide range of $l_y$ distributed over the surface.
A noteworthy feature is that very tall structures with heights approaching $\delta$ are more evident in internal flows (figure \ref{fig:1}$b$,$c$) compared with TBL (figure \ref{fig:1}$a$). 
These large structures meander in the spanwise direction \citep{Hutchins07} and appear with the streamwise length over 10$\delta$ in internal flows \citep{Monty07,Monty09}.
Such a difference in very long structures may imply that we need to filter out those structures to examine the self-similar features of energy-containing motions in the logarithmic region.

\begin{figure}
\centerline{\includegraphics[trim=0cm 0.2cm 0cm 0.2cm,width=14cm]{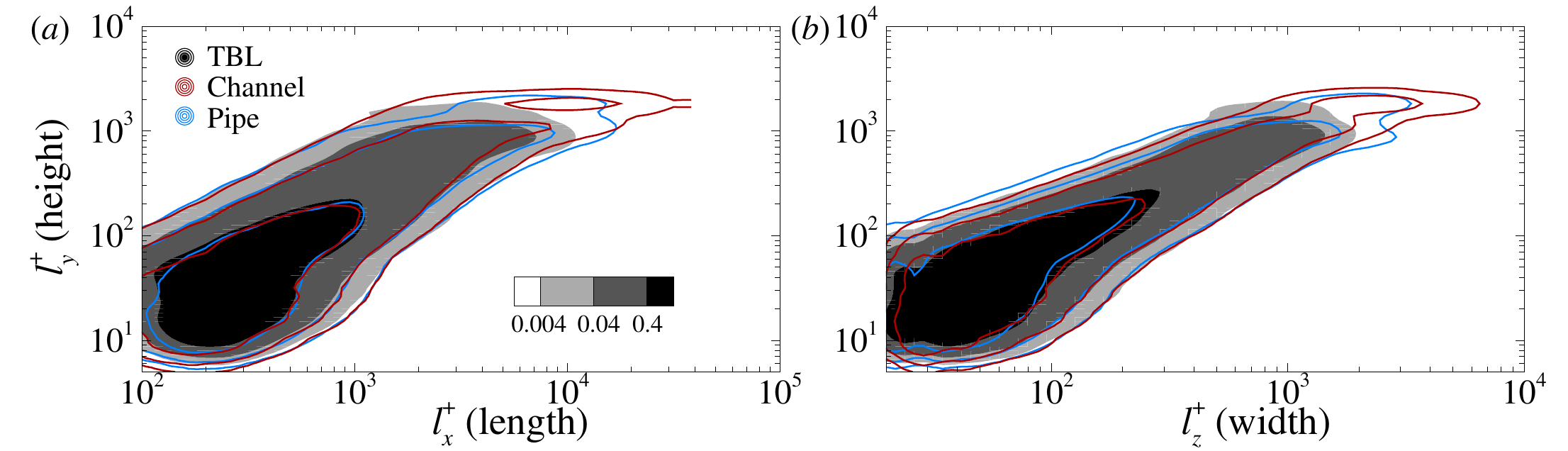}}
\caption{\label{fig:2} ($a$,$b$) Joint probability density functions (PDFs) of the logarithms of the streamwise length ($l_x$) and spanwise width ($l_z$) of wall-attached structures with respect to their wall-normal heights ($l_y$). Here, the shaded, red line, and blue line contours indicate the datasets of the TBL, channel, and pipe, respectively. Contour levels are logarithmically distributed.}
\end{figure}

To quantify the geometrical features of wall-attached structures, the joint PDFs of the sizes of the identified structures are plotted in figure \ref{fig:2}.
Here, the shaded, red line, and blue line contours denote the datasets of the TBL, channel, and pipe flows, respectively.
In general, the length ($l_x$) and width ($l_z$) of the structures increase with increasing heights ($l_y$) in all three flows.
Interestingly, the distributions of $l_x$ and $l_z$ show a reasonably good collapse over a wide range of $l_y$ except for $l_y \approx O(\delta)$.
This result indicates that the geometric features of the wall-attached $u$ structures in the inner region are presumably universal in canonical wall turbulence.
On the other hand, there are very-large-scale structures with $l_y \approx O(\delta)$, and some with $l_y \approx 2\delta$, in internal flows (red and blue line contours), which is consistent with the largest clusters of ejections or sweeps in channel flows \citep{Lozano12}.
The sizes of these structures are analogous to those of very-large-scale motions \citep{Kim99} or global modes \citep{Del04} reported in internal flows.
In TBL, very tall structures ($l_y \approx O(\delta)$) are associated with superstructures \citep{Hutchins07} because $l_x$ extends over $6\delta$ (see figure \ref{fig:1}$a$) and they are physically attached to the wall (i.e. footprints in the near-wall region).
Recent studies \citep{Hwang18,Baars20,Yoon20} have reported that tall wall-attached structures are non-geometrically-similar (or non-self-similar).
In other words, this may reflect the fact that the statistical characteristics of these motions are non-universal and can be directly affected by the flow geometry.
Hence, we need to filter out tall wall-attached structures in order to analyse the existence of self-similar features in the logarithmic layer \jya{\citep{Hwang18,Baars20}}.

\begin{figure}
\centerline{\includegraphics[trim=0cm 0.2cm 0cm 0.2cm,width=14cm]{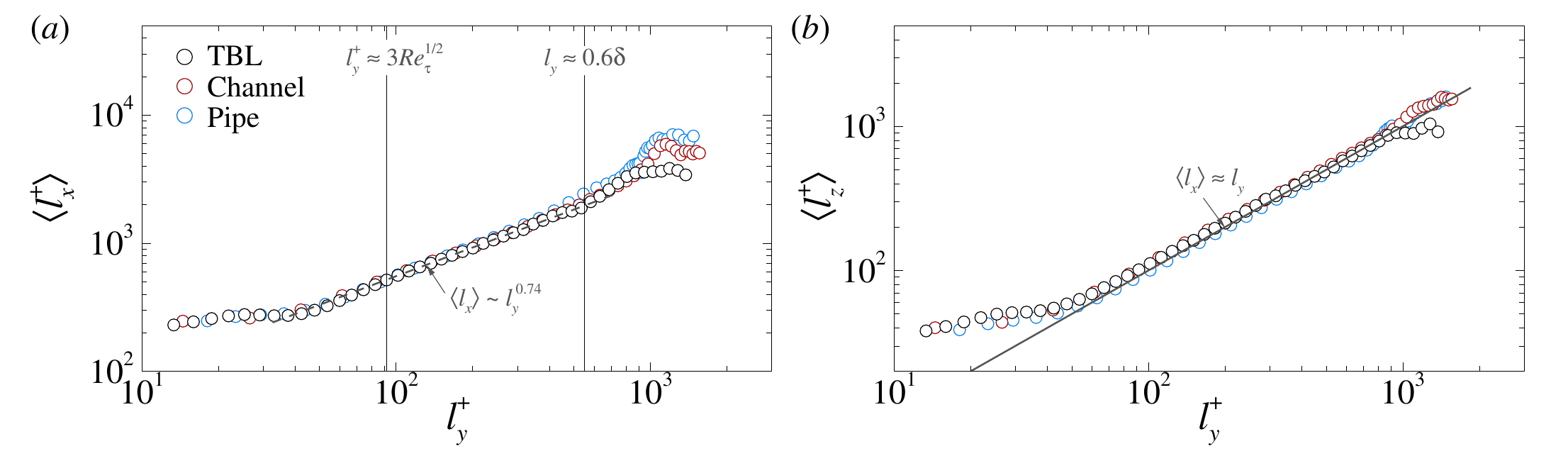}}
\caption{\label{fig:3} Mean length ($\langle l_x \rangle$) and width ($\langle l_z \rangle$) of wall-attached structures as a function of $l_y$. In ($a$), the dashed line is $\langle l_x \rangle \sim l_y^{0.74}$ \citep{Hwang18}. In ($b$), the solid line corresponds to $\langle l_z \rangle \approx l_y$.}
\end{figure}

Figure \ref{fig:3} presents the variation of the mean streawmwise length $\langle l_x \rangle$ and the mean spanwise width $\langle l_z \rangle$ at a given $l_y$.
The growth rates of $\langle l_x \rangle$ and $\langle l_z \rangle$ with respect to $l_y$ are comparable in all three flows. 
In figure \ref{fig:3}($b$), we can see the linear relationship \jya{$\langle l_z \rangle \sim l_y$} over $l_y^+ > 3Re_\tau^{1/2}$.
In contrast, $\langle l_x \rangle$ shows a power-law behaviour (i.e. $\langle l_x \rangle \sim l_y^{0.74}$) over \jyb{$3Re_\tau^{1/2} < {l_y}^+ < 0.6\delta^+$} (figure \ref{fig:3}$a$)  in all three flows.
For \jyb{$l_y^+ \ge 0.6\delta^+$}, the distribution of $\langle l_x \rangle$ starts to deviate away from the power law ($\langle l_x \rangle \sim l_y^{0.74}$).
Although there is no linear relationship for the wall-attached $u$ structures over $3Re_\tau^{1/2} \le l_y^+ < 0.6\delta^+$ in figure \ref{fig:3}($a$), the sizes of the structures are scaled with $l_y$ and show a good agreement regardless of flow geometry.
This result supports the supposition that the wall-attached structures in this range may share similar features.
Moreover, \jya{the mean length $\langle l_x \rangle$ is approximately $3\delta$ at the upper limit $l_y^+ = 0.6\delta^+$, which is consistent with the criteria to distinguish between large-scale and very-large-scale motions \citep{Guala06,Balakumar07,Wu12,Hwang16b,Hwang16}.}
This limit also corresponds to the criteria that distinguish self-similar structures from non-self-similar ones in an adverse-pressure-gradient TBL \cite{Yoon20}.
\jya{Notably, $l_y^+ = 0.6\delta^+$ is larger than the wall-normal location of the outer peak in the one-dimensional premultiplied spectra of $u$, which characterises very-large-scale motions or superstructures.
Given that $l_y$ is measured from the bounding box of each structure, the wall-parallel area of each structure is close to zero as $y$ approaches $l_y$, and they have the maximum wall-parallel area at $y < l_y$.}

Hence, the wall-attached $u$ structures can be classified into three components in terms of $l_y$; buffer-layer, self-similar, and non-self-similar structures defined as $l_y^+ < 3Re_\tau^{1/2}$, $3Re_\tau^{1/2} \le l_y^+ < 0.6\delta^+$, and $0.6\delta \le l_y^+$, respectively.
Here, the lower bound ($= 3Re_\tau^{1/2}$ of self-similar structures corresponds to that of the logarithmic region in \cite{Marusic13,Hwang19}, which is a mesolayer scaling \citep{Afzal82,Wei05}.
Given that the viscous effect may affect the coherent motions in the logarithmic region \citep{Hwangy16}, the bound $3Re_\tau^{1/2}$ is used in the present study.
The present Reynolds number is $Re_\tau \approx 1000$ in which, in turn, the lower bound is approximately 100 wall units (i.e. $3Re_\tau^{1/2} \approx 100$), that is, a classical scaling for the lower bound of the logarithmic region \citep{Perry82}.
Since we focus on the logarithmic region where $y^+ > 100$, the variation of the lower bounds does not have a major impact on our conclusions.

\jya{The power law ($\langle l_x \rangle \sim l_y^{0.74}$) for the WASS could be attributed to the low Reynolds number of the present data.
At $Re_\tau \approx 3000$, the linear relationship ($\langle l_x \rangle \sim l_y$) was observed for $l_y^+ > 400$ in a turbulent pipe flow \citep{Hwang19}.
Given $\langle l_z \rangle \sim l_y$, the wall-attached structures with $3Re_\tau^{1/2} \le l_y^+ < 0.6\delta^+$ follow $\langle l_x \rangle l_y \sim \langle l_z \rangle^{1.74}$ , which is roughly quadratic.
This behaviour is consistent with the scaling of the two-dimensional spectra of $u$ at $Re_\tau < 2000$\citep {Del04} in which the energetic ridges of the spectra are aligned along $\lambda_x y \sim \lambda_z^2$ in the logarithmic region.
\cite{Chandran17} reported that the two-dimensional spectra at low Reynolds number follow $\lambda_x y \sim \lambda_z^2$ whereas the larger scales tend to exhibit the linear law ($\lambda_x \sim \lambda_z$) at high Reynolds numbers ($Re_\tau = 26000$).
Although we defined the length scales of wall-attached structures based on the sizes ($l_x, l_y$, and $l_z$) of the bounding box of each structure, this result supports the inference that $l_x, l_y$, and $l_z$ can be used to describe the characteristic length scales of the energetic motions in the logarithmic region.
It should be noted that these scales are measured from the physically connected volumes of intense $u$, which consist of the energy contributions from a wide range of scales (or wavelengths).
In \S3.2, the two-dimensional energy spectra are examined to elucidate the wavelength scales contained within the WASS.}

\begin{figure}
\centerline{\includegraphics[trim=0cm 0.2cm 0cm 0.2cm,width=14cm]{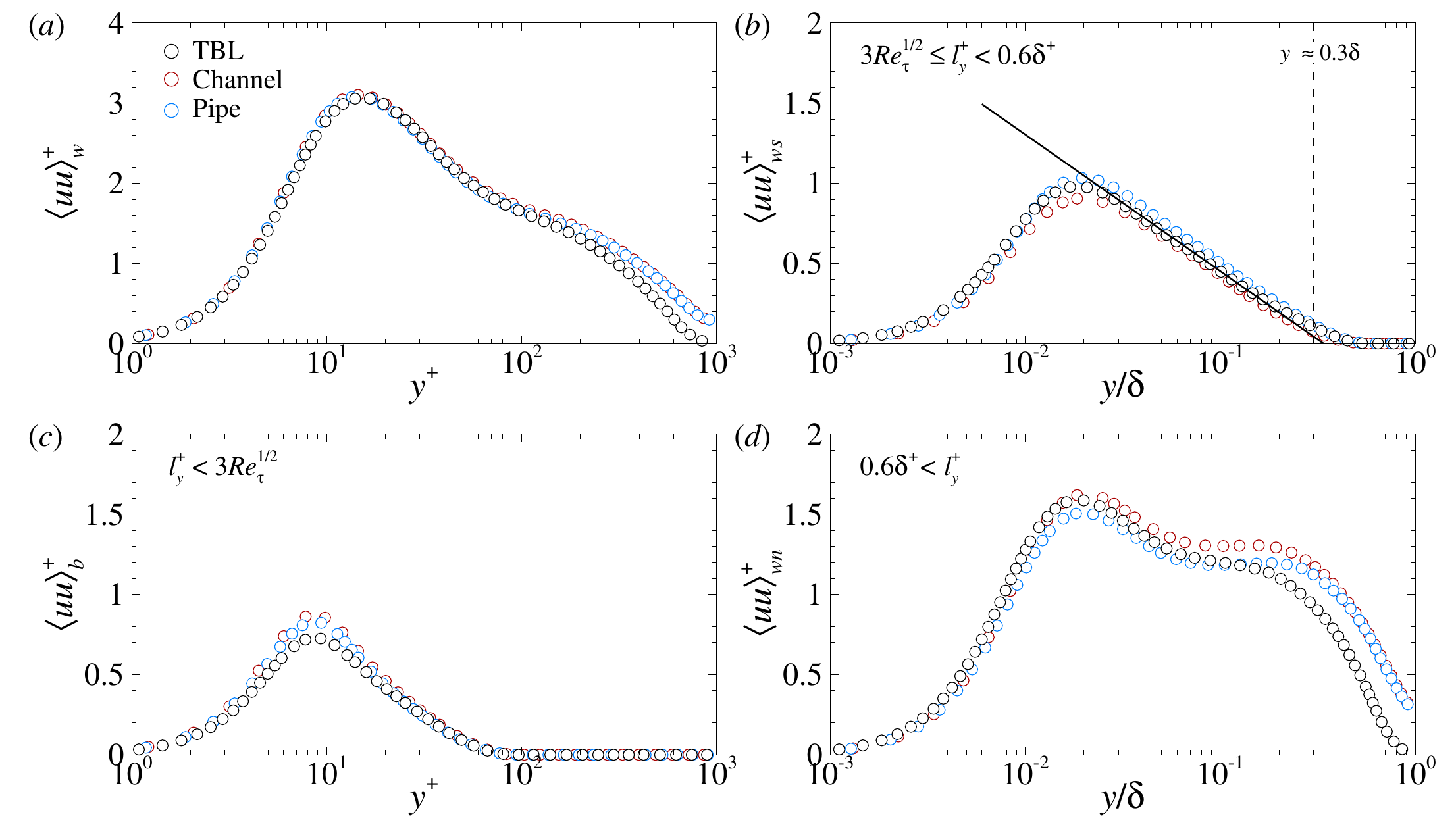}}
\caption{\label{fig:4} Wall-normal profiles of streamwise turbulence intensity carried by wall-attached structures: ($a$) $\langle uu \rangle_w^+$; ($b$) $\langle uu \rangle_{ws}^+$; ($c$) $\langle uu \rangle_{b}^+$; and ($d$) $\langle uu \rangle_{wn}^+$. \jya{In ($b$), the solid line denotes a logarithmic variation corresponding to $\langle uu \rangle_{ws}^+ = 0.4 - 0.37\ln(y/\delta)$.}}
\end{figure}

To test the reliability of the present decomposition, we compute the streamwise turbulence intensity distributed among the three components of wall-attached $u$ structures.
The streamwise velocity fluctuations associated with each component ($u_i$) can be conditionally sampled based on the bounded volume of each object:
\begin{equation}
\begin{aligned}
  u_{b}(\bold{x}) &= \left\{
    \begin{array}{ll}
      u(\bold{x}), & \mbox{if\ } \bold{x} \in \Omega_b, \\[2pt]
      0, & \mbox{otherwise},
  \end{array} \right.
 \\
  u_{ws}(\bold{x}) &= \left\{
    \begin{array}{ll}
      u(\bold{x}), & \mbox{if\ } \bold{x} \in \Omega_{ws}, \\[2pt]
      0, & \mbox{otherwise},
  \end{array} \right.  
  \\
  u_{wn}(\bold{x}) &= \left\{
    \begin{array}{ll}
      u(\bold{x}), & \mbox{if\ } \bold{x} \in \Omega_{ws}, \\[2pt]
      0, & \mbox{otherwise},
  \end{array} \right.    
  \label{eq:eq2}
\end{aligned}
\end{equation}
where $\Omega_i$ denotes the positions of \jya{all contiguous points of each object}.
The subscripts $i$ \textit{b}, \textit{ws}, and \textit{wn} refer to wall-attached buffer-layer ($l_y^+ < 3Re_\tau^{1/2}$), self-similar ($3Re_\tau^{1/2} \le l_y^+ < 0.6\delta^+$), and non-self-similar structures ($0.6\delta \le l_y^+$), respectively.
This classification yields a computation of the streamwise turbulence intensity corresponding to the specific structure.
For example, the streamwise turbulence intensity carried by wall-attached self-similar structures (WASS) $\langle uu \rangle_{ws}$ can be obtained through the ensemble average of $u_{ws}u_{ws}$ at a given wall-normal position.

Figure \ref{fig:4} shows the wall-normal distributions of the streamwise turbulence intensity carried by the specific motion.
\jya{Notably, we can observe the presence of the near-wall peak in all streamwise turbulence intensities because the identified structures physically adhere to the wall.
Hence, the near-wall part (or footprint) of each object is related to the near-wall streaks \citep{Kline67}, reflecting that the identified structures could be responsible for the near-wall turbulence.
However, this is beyond the scope of the present work, which focuses on the logarithmic region.}
The total streamwise turbulence intensity ($\langle uu \rangle_{w}$) contained within all three components is shown in figure \ref{fig:4}($a$).
Here, $\langle uu \rangle_{w}$ is obtained using the streamwise velocity fluctuations within the wall-attached structures ($= u_b + u_{ws} + u_{wn}$).
As seen, there is a complete collapse of the wall-normal profiles of $\langle uu \rangle_{w}$ in internal flows across the wall-normal direction.
In contrast, the $\langle uu \rangle_{w}$ of TBL shows a discrepancy above the logarithmic region ($y^+ > 100$).
This discrepancy occurs at a lower wall-normal location compared with the results of \cite{Monty09}, who reported that the difference in $\langle uu \rangle$ between internal and external flows appears at $y > 0.5\delta$.
Given that $\langle uu \rangle_{w}$ represents the collective contributions of the wall-attached structures, the deviation in figure \ref{fig:4}($a$) is due to the structures with $l_y^+ > 100$, which include WASS and wall-attached non-self-similar structures (WANS).
\jya{The weak-$u$ region ($|u| < u_{rms}$ ) also significantly contributes to the total streamwise turbulence intensity because this region has a larger area fraction than the intense $u$ region (\ref{eq:eq1}).
The area fraction of the wall-attached $u$ structures is only 6--7\% in the logarithmic region.
However, the contribution of the wall-attached $u$ structures to the total turbulence intensity is approximately 40\%, which is comparable to that of the weak-$u$ region.
This may lead to the absence of the logarithmic variation in the total streamwise turbulence intensity at the present $Re_\tau$ \citep{Hwang18}.}

\jya{Figure \ref{fig:4}($b$)} displays the streamwise turbulence intensity carried by the WASS ($\langle uu \rangle_{ws}$).
In all three flows, the profiles of $\langle uu \rangle_{ws}$ show good agreement (figure \ref{fig:4}$b$), indicating that the WASS are independent of the flow geometry \citep{Perry86}. 
In addition, there is a logarithmic variation of up to $y = 0.3\delta$ in all three flows, reminiscent of Townsend's attached eddies.
In figure \ref{fig:4}($c$), the streamwise turbulence intensity of the buffer-layer structures $\langle uu \rangle_b$ also exhibits a reasonable match in all three flows.
Hence, the similarity in the streamwise turbulence intensity in all three flows (figure \ref{fig:4}$b$,$c$) suggests that geometrical differences are negligible in the near-wall region.

In contrast, figure \ref{fig:4}($d$) shows that the $\langle uu \rangle_{wn}$ of internal flows appear to be larger than that of TBL in the outer region.
This supports that the discrepancy in $\langle uu \rangle_w$ over the logarithmic region (figure \ref{fig:4}$a$) originates from the non-self-similar nature of very tall structures (i.e. WANS), which extend from the near-wall region to the core region of internal flows.
In addition, comparing pipe and channel data, $\langle uu \rangle_{wn}$ of the channel flow is larger than that of the pipe flow in the logarithmic region. 
This might be attributable to a larger population of very long $u$ streaks in the channel flow \citep{Lee15}.
Owing to pipe curvature, the sizes of very large structures could be restricted in pipe flows; this leads to the dominant contributions of large scales in the channel flow \citep{Hwang16b,Han19}.

\jya{It is worth mentioning that the magnitude of the slope of the logarithmic variation in figure \ref{fig:4}($b$) is 0.37, which is lower than the Townsend--Perry constant reported in previous studies (1.26 by \cite{Marusic13}, 0.98 by \cite{Baars20b}, and 0.8--1.0 by \cite{Hu20}).
This is due to the decomposition method used here in which $u_{ws}$ is zero at the outside of the identified structures (\ref{eq:eq2}).
As a result, the magnitude of $\langle uu \rangle_{ws}$ decreases, which leads to a lower slope of the logarithmic variation.
However, we primarily focus on the presence of the logarithmic variation in streamwise turbulence intensity and the collapse of the corresponding profiles in all three flows obtained from the classification of wall-attached $u$ structures in terms of height.
The presence of the logarithmic variation could be a consequence of the spatial organisation (self-similarity and distributions) of WASS in the context of the AEH.}

The present decomposition method filters wall-attached structures that are geometrically self-similar.
By filtering out the contributions of WANS in the logarithmic region, we can analyse the spectral signature of self-similar coherent motions.
In the next section, the two-dimensional energy spectra of $u_{ws}$ are examined by focusing on the self-similar scaling laws proposed by Perry and coworkers \citep{Perry77,Perry82,Perry86}, and on the energy distribution across the streamwise and spanwise wavelengths ($\lambda_x$ and $\lambda_z$) in the logarithmic region.

\subsection{Two-dimensional spectra of wall-attached self-similar structures}\label{sec:2D}
\begin{figure}
\centerline{\includegraphics[trim=0cm 0.2cm 0cm 0.2cm,width=14cm]{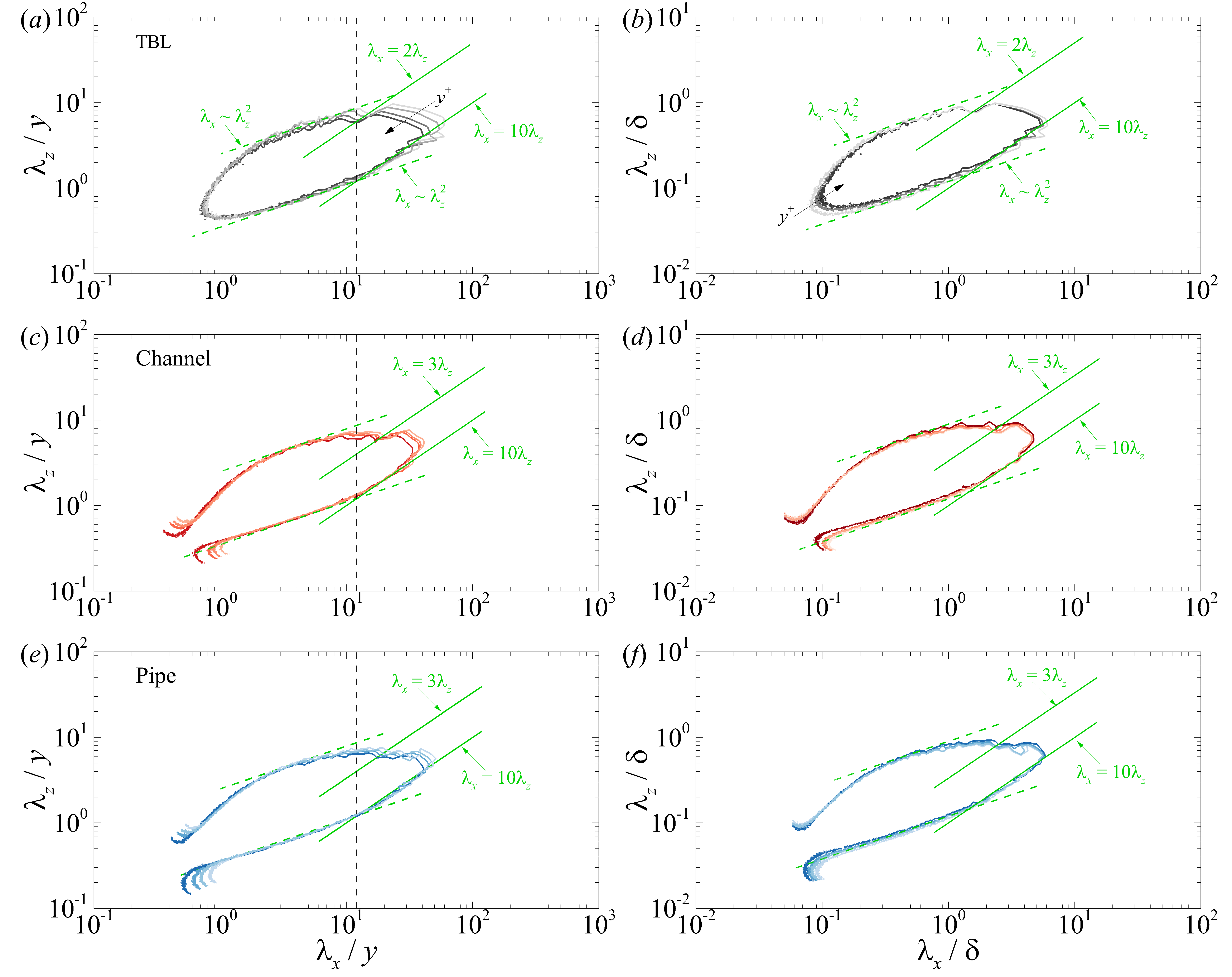}}
\caption{\label{fig:5} Premultiplied two-dimensional energy spectra $\Phi^{2D}$ across the logarithmic region ($y^+ = 100, 110, 120$, and $130$): ($a,b$) TBL; ($c,d$) channel; and ($e,f$) pipe. Light to dark shading indicates an increase in $y$. The contour level is $0.4$ times each of the maxima. Here $\Phi^{2D}$ is plotted as a function of the wall-parallel wavelengths ($\lambda_x$ and $\lambda_z$) normalised by $y$ ($a$,$c$,$e$) and $\delta$ ($b$,$d$,$f$). The green dashed and solid lines denote $\lambda_x \sim \lambda_z^2$ and $\lambda_x \sim \lambda_z$, respectively. The vertical dashed line indicates $\lambda_x = 12y$.}
\end{figure}

The contribution of coherent motions ingrained within WASS to the streamwise variance is explored by computing the two-dimensional spectrum.
The premultiplied two-dimensional spectrum of $u_{ws}$, $\Phi^{2D}$. is defined as
\begin{equation}
\Phi^{2D}(k_x, k_z, y) = k_x k_z \langle  \hat{u}_{ws}(k_x, k_z, y) \hat{u}^{*}_{ws}(k_x, k_z, y) \rangle,
  \label{eq:eq3}
\end{equation}
where $k_x (= 2\pi/\lambda_x)$ and $k_z (= 2\pi/\lambda_z)$ are the streamwise and spanwise wavenumbers and $\hat{u}_{ws}$ indicates the Fourier coefficient of $u_{ws}$ and the asterisk denotes a complex conjugate.
The streamwise variance of WASS ($\langle uu \rangle_{ws}$) is the integral of the corresponding two-dimensional spectrum over the wall-parallel wavenumbers.
\jya{Notably, $l_x$ and $l_z$ of the identified structures, measured from the bounding box, do not necessarily represent the characteristic scales at a given wall-parallel plane ($y < l_y$) because the structures are inclined with respect to the wall and meander in the spanwise direction.
Given that $u_{ws}$ represents the conditionally sampled $u$ which are contained within the WASS with $y \le l_y < 0.6\delta$, $\Phi^{2D}$ could provide the energy distributions from the collective contribution of WASS.}
In other words, although the ranges of the corresponding physical length and width are 5--6$y < \langle l_x \rangle < 2\delta$ and $y < \langle l_z \rangle < 0.6\delta$ (figure \ref{fig:3}), the energy in $\Phi^{2D}$ is distributed across a wide range of wavelengths.
\jya{This also supports the inference that the structures identified in the physical space consist of a broad range of turbulent motions related to the concept of nested hierarchies \citep{Hwang18}.}
Here, the energy contained in the very long and wide wavelengths ($\lambda_x > 6\delta$ and $\lambda_z > \delta$), which correspond to global modes \citep{Del04,Jimenez08}, is negligible because the velocity fluctuations associated with WANS are subtracted.
Hence, we can expect that the energy distribution demonstrates the spectral signature of self-similar coherent motions.
\jya{The effect of the threshold $\alpha$ was examined and the results reported in this section remained qualitatively unchanged (see appendix \ref{appA}).}

Figure \ref{fig:5} shows the contours of $\Phi^{2D}$ at various wall-normal positions located in the logarithmic region ($y^+ = 100, 110, 120$, and $130$) for all three flows.
Here, $\Phi^{2D}$ as a function of the wavelengths scaled with $y$ and $\delta$ is plotted on the left and right columns, respectively.
In general, all three flows show that $\Phi^{2D}$ scales reasonably well with $y$ in the range $y < \lambda_x < 10y$ (figure \ref{fig:5}$a,c,e$) and with $\delta$ in the range $\lambda_x > \delta$ (figure \ref{fig:5}$b,d,f$).
To further examine the spectral behaviour, the bounds of the constant energy distribution are denoted by the green lines.
Here, the dashed and solid lines represent a square-root $\lambda_x \sim \lambda_z^2$ and linear relationship $\lambda_x \sim \lambda_z$, respectively.
The contours of $\Phi^{2D}$ in figure \ref{fig:5}($a,c,e$) are bounded by \jya{$\lambda_x \sim \lambda_z^2$} over the range $y < \lambda_x < 10y$. 
This is consistent with the results of the two-dimensional spectra of $u$, which include all coexisting motions, at low $Re_\tau \approx O(10^3)$ in the works of \cite{Del04} and \cite{Chandran17}.
According to \cite{Del04} and \cite{Chandran17}, the linear relationship was absent in the large-scale range at low $Re_\tau$.
In the large-scale range ($\lambda_x > 12y$), however, $\Phi^{2D}$ are roughly aligned along $\lambda_x \sim \lambda_z$ (solid lines), which is an indicator of self-similarity.
Moreover, the contour lines of $\Phi^{2D}$ collapse reasonably well along the lower bound $\lambda_x = 10\lambda_z $ with both $y$ and $\delta$ scaling in this large-scale range, reflecting the overlap arguments of self-similar energy-containing motions \citep{Perry77,Perry82,Perry86}.
This result also supports the existence of Townsend's attached eddies, even in low $Re_\tau$ \citep{Hwang18}, and further reveals that the WASS, defined in physical space, is tightly connected to the spectral signatures of the AEH in the framework used by Perry and coworkers.

It is worth noting that the linear relationship of the upper bound in the large-scale range ($\lambda_x \approx 2-3\lambda_z$) seems to appear over a small range owing to low $Re_{\tau}$.
At high $Re_\tau = O(10^4)$, the upper and lower bounds of the two-dimensional spectra follow the linear relationship in the large-scale range \citep{Chandran17,Deshpande20}.
As shown in figure \ref{fig:5}($b,d,f$), the contour lines of $\Phi^{2D}$ in the large scale range are aligned along the horizontal line $\lambda_z \approx 0.8\delta$.
Given that the growth of energy-containing motions can be restricted as their heights reach $\delta$, there may not be enough space to maintain large-scale self-similar motions at the present Reynolds number.
According to the AEH, self-similar energy-containing motions fill in the scale separation between $\nu/u_\tau$ and $\delta$.
At high Reynolds numbers, it can be conjectured that the contours of $\Phi^{2D}$ in the outer scaling would move to the bottom left corner and lead to a clear upper bound that follows the linear behaviour.
This upper bound may lie below $\lambda_x = \lambda_z$ (i.e. a higher slope than unity) due to the inclination nature of wall-attached $u$ structures (figures \ref{fig:1} and \ref{fig:3}$a$).
In other words, it may imply that self-similar behaviour can be observed when large-scale energy-containing motions have a high aspect ratio $\lambda_x/\lambda_z$ \citep{Chandran17}.

As discussed, the linear behaviour of the lower bound ($\lambda_x = 10\lambda_z$) is observed in the large-scale range where $\lambda_x > 12y$ (denoted by vertical dashed lines in figure \ref{fig:5}$a,c,e$).
Such a feature is found in all three flows indicative of the universality of the self-similar nature of the large scales contained within WASS.
In addition, $\lambda_x > 12y$ is consistent with the lower limit of the $k_x^{-1}$ region in channel flows \citep{Hwangy15}, and is close to the inner-scaling limit ($\lambda_x = 14y$) of self-similar motions in TBLs over a wide range of $Re_\tau$ \citep{Baars17}.
A recent study by \cite{Deshpande20} also found the existence of linear behaviour in two-dimensional cross spectra of wall-coherent $u$ motions over a similar range (i.e. $\lambda_x > 15y$).
This shows that the WASS identified in the present study is directly related to the self-similar behaviour in the wavenumber space. 
In \S\ref{sec:1D}, the one-dimensional spectra of WASS is examined to observe the presence of the $k_x^{-1}$ region over a similar large-scale range ($\lambda_x > 12y$), which was absent in the aforementioned study.

\jya{It is worth mentioning that $u_{ws}$ could demarcate the rapid change in the raw $u$.
Wall-attached $u$ structures are composed of multiple UMZs \citep{Hwang18}.
Given that the UMZs are demarcated by a thin shear layer or large velocity gradient \citep{Meinhart95, Adrian00}, the boundaries of the identified structures may be one of the internal shear layers where the streamwise momentum exhibits a sharp change in the velocity.
We can observe a rapid variation of the raw velocity signals near the edges of the identified structure; see figure 9 in \cite{Hwang18}.
This result supports the inference that $u_{ws}$ can retain the spectral characteristics of the raw $u$ signals even if we artificially impose zero velocity outside of the identified structures.
\cite{Srinath18} also reported that binary representation of the negative-$u$ structures could preserve the spectral information of raw $u$ because of large changes in $u$ at the edges of the identified structures.}

\begin{figure}
\centerline{\includegraphics[trim=0cm 0.2cm 0cm 0.2cm,width=14cm]{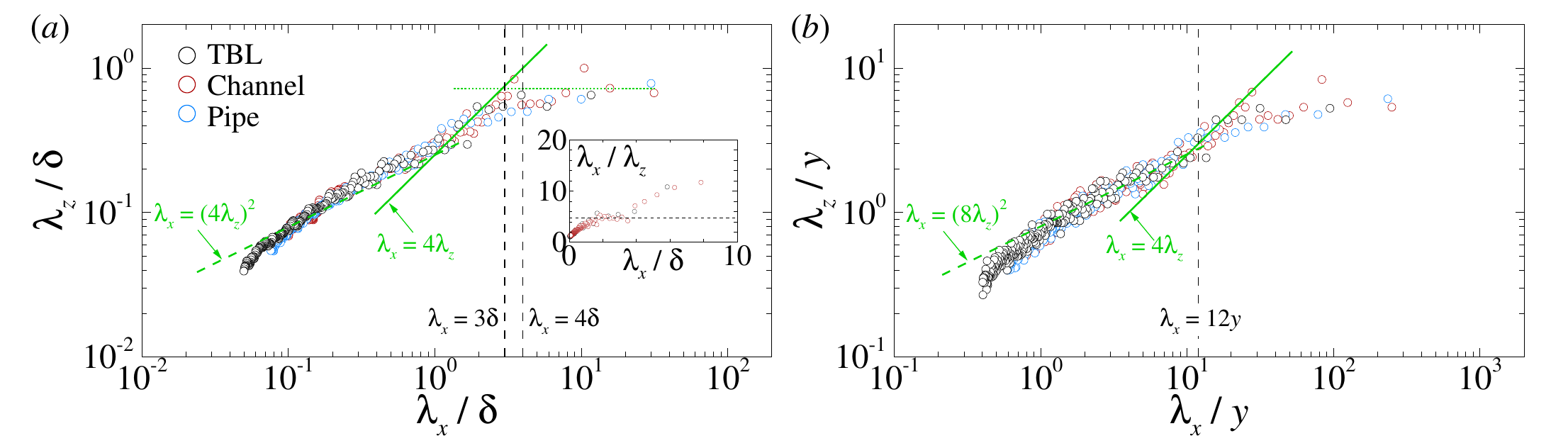}}
\caption{\label{fig:6} Energetic ridges of the premultiplied two-dimensional energy spectra $\Phi^{2D}$ at $y^+ = 120$. Here, the positions of the ridges are obtained by identifying $\lambda_z$ of the maximum $\Phi$ at a given $\lambda_x$. The dashed and solid green lines represent the power-law ($\lambda_x = (4\lambda_z)^2$) and linear relationships ($\lambda_x = 4\lambda_z$), respectively.
\jya{In ($a$), the inset shows the lin-lin plot of the ridge scale ratio $\lambda_x / \lambda_z$ for the TBL and channel data. The horizontal dashed line denotes a constant ratio $\lambda_x / \lambda_z \approx 4.$}}
\end{figure}

To further explore the energy contribution from self-similar motions, the spectral ridge of $\Phi^{2D}$ is plotted in figure \ref{fig:6}.
Here, the spectral ridge is determined by identifying $\lambda_z$ corresponding to the maximum value of $\Phi^{2D}$ at a given $\lambda_x$.
Hence, figure \ref{fig:6} represents the length-scale ($\lambda_x$ and $\lambda_z$) relationship of the energetic motions.
All of the spectral ridges are found to agree reasonably well over a wide range of scales.
We can observe two growth rates: one with the power-law behaviour $\lambda_x \sim \lambda_z^2$ (dashed green line), and one with the linear relationship $\lambda_x \sim \lambda_z$ (solid green line) at relatively large scales.
In addition, the transition of the ridges from the power law to the linear law appears at $\lambda_x = 12y$.
This result is consistent with the variation of the lower and upper bounds of $\Phi^{2D}$ found in figure \ref{fig:5} and reflects that the dimensions of the energetic motions behave in an analogous manner to the wavelength relationship of the bounds (i.e. self-similarity).
It is noted that the spectral ridges flatten for $\lambda_x > 3-4\delta$ because at this very long $\lambda_x$ the spanwise wavelength $\lambda_z$ reaches $\delta$, and the growth of $\lambda_z$ is restricted.
\jya{A similar spectral trend was observed in \cite{Chandran17} and \cite{Deshpande20}.
Given that the wall-attached $u$ structures follow $\langle l_z \rangle \approx l_y$ (figure \ref{fig:3}$b$), it may reflect the saturation in their spanwise growth.}
Although the motions related to these scales are non-self-similar, the energy contribution from the range $\lambda_x > 3-4\delta$ and $\lambda_z > 0.8\delta$ is negligible, as shown in figure \ref{fig:5}, since $\Phi^{2D}$ is obtained from $u_{ws}$ (\ref{eq:eq2}).

In figure \ref{fig:6}, the linear behaviour seems to follow $\lambda_x \approx 4\lambda_z$ over the range $12y < \lambda_x < 3-4\delta$.
\jya{Here, the inset shows the ridge scale ratio $\lambda_x / \lambda_z$ for the TBL and channel data.
The data over this range are not very far from the dashed line ($\lambda_x / \lambda_z \approx 4$). }
Hence, the range of the self-similar energetic motions ingrained in WASS can be expressed by
\begin{subeqnarray}
12y & < & \lambda_x < 3-4\delta,\\[3pt]
3y & < & \lambda_z < 0.8-1\delta.
  \label{eq:eq4}
\end{subeqnarray}
Here, the upper limit for the streamwise wavelength in (\ref{eq:eq4}$a$) is $3-4\delta$ similar to that of the criteria that distinguishes large-scale motions and very-large-scale motions \citep{Guala06, Balakumar07,Wu12,Hwang16b}.
In particular, the upper limit of the TBL ($\approx 3\delta$) is relatively smaller than that of the internal flows ($\approx 4\delta$), which may be related to the dominant contribution of very long scales in internal flows \citep{Monty09}.
Given that the contours of $\Phi^{2D}$ are restricted to below $\lambda_x \approx 3-4\delta$ in figure \ref{fig:5}, this result also supports that $\Phi^{2D}$ is composed of the contribution from the turbulence motions that include large-scale motions and relatively smaller motions.

It is worth mentioning that a similar linear relationship of wall-attached $u$ structures was found by \cite{Hwang19}, who showed that the physical length and width of WASS exhibit $\langle l_x \rangle = \langle 4l_z \rangle$ in a higher Reynolds number pipe flow ($Re_\tau \approx 3000$).
This result may reflect that, although there is no linear behaviour in the physical length ($\langle l_x \rangle$) of WASS (figure \ref{fig:2}), the large-scale motions ingrained in WASS are self-similar at a lower $Re_\tau$. 
In addition, these motions become prominent in the physical space when there is enough space in the logarithmic region caused by the length scale separation.
According to \cite{Deshpande20}, the two-dimensional spectra of wall-coherent motions at $Re_\tau \approx 15000$ are aligned along a linear ridge $\lambda_x = 7 \lambda_z$, which is slightly steeper compared with the proportionality $\lambda_x = 4\lambda_z$ found in the present work.
However, the spectra reported by \cite{Deshpande20} includes the contributions from both wall-attached self-similar and non-self-similar motions.
Given that very large scales can contaminate self-similar behaviours of turbulent motions \citep{Jimenez08,Hwang18,Han19}, it would be instructive in future efforts to examine the energetic ridges of $\Phi^{2D}$ over a wide range of $Re_\tau$.

The two distinct ridges might reflect a bimodal behaviour of self-similar energy-containing motions.
In turbulent channel flows, \cite{Hwangy15} found that the energy-containing motions consist of two distinct motions, of which one is related to long streaky motions ($\lambda_x \approx 10\lambda_z$) and the other is associated with packets or clusters of vortical motions ($\lambda_x \approx 2-3\lambda_z$).
The aspect ratio ($\lambda_x/\lambda_z$) of the latter motion carrying all velocity components of the turbulent kinetic energy is approximately similar to that of the WASS in the present work.
This also agrees with the dimensions of tall vortex clusters \citep{Del06b} and tall ejection/sweep clusters \citep{Lozano12}; here, the use of `tall' denotes structures that extend beyond the logarithmic region.
Owing to low $Re_{\tau}$ in the present work, the growth of self-similar motions was restricted by (\ref{eq:eq4}), which in turn leads to a lower contribution of self-similar motions with a high aspect ratio in the logarithmic region.
As shown in figure \ref{fig:5}, the lower bound of $\Phi^2D$ is aligned along $\lambda_x = 10 \lambda_z$ and presents the $y$ and $\delta$ scalings simultaneously.
The high aspect ratio motions are dominantly ingrained in WANS at the present $Re_\tau$ and thus we may not observe the self-similarity of the high aspect ratio motions ($\lambda_x/\lambda_z = 7$) in the range ($\ref{eq:eq4}a$).

\subsection{One-dimensional spectra of wall-attached self-similar structures}\label{sec:1D}
\jya{In this section, we examine the one-dimensional spectra to further analyse the self-similar scaling observed in the two-dimensional spectra}.
According to Perry and coworkers \citep{Perry77,Perry82,Perry86}, the $k_x^{-1}$ scaling in the one-dimensional spectra of $u$ can serve as a spectral signature of energy-containing motions satisfying the AEH.
Although \cite{Nickels05} reported the presence of the $k_x^{-1}$ region, such a scaling has remained ambiguous at extremely high $Re_\tau$ \citep{Rosenberg13}.
It seems that the $k_x^{-1}$ region may exist when the one-dimensional spectra are plotted in the log$-$log form \citep{Vallikivi15}, but the log$-$log plot of the one-dimensional spectra without premultiplication exaggerates the $k_x^{-1}$ scaling \citep{Baars20}.
The ambiguity of such a scaling law is also related to the aliasing in the one-dimensional spectra because the one-dimensional streamwise spectra are a measure of average energy distribution without including the spanwise information; thus, the energy carried by smaller wavenumbers (i.e. large scales) can be contaminated \citep{Tennekes72,Davidson06}.
Given the fact that self-similar behaviours appear at large sclaes, this supports the claim that the $k_x^{-1}$ law is relatively hard to observe even when we can observe the logarithmic variation in the streamwise turbulence intensity \citep{Hultmark12,Marusic13}.
\jya{It is worth noting that \cite{Srinath18} showed the relation between the wall-attached $u$ structures and one-dimensional spectra, i.e. the one-dimensional spectra follow a $k_x^{-1-p}$ scaling in which the exponent $p$ is related to the streamwise turbulence intensity within the identified structures.
Although they found that the exponent $p$ becomes zero (i.e. $k_x^{-1}$ scaling) over $y^+ = 100$--200, it has not been revealed whether this spectral behaviour is associated with the self-similar scaling in the two-dimensional spectra.}

The analysis of the two-dimensional spectra can avoid the aliasing issue.
By analysing the two-dimensional spectra, \cite{Chandran17} and \cite{Deshpande20} successfully showed the existence of self-similar turbulent motions in the logarithmic region.
However, there is no clear $k_x^{-1}$ region over the range where the self-similarity appears in the two-dimensional spectra, although \cite{Deshpande20} obtained the spectra of wall-coherent motions of $u$.
The absence of the $k_x^{-1}$ region presumably originates from coexisting scales that are non-self-similar \citep{Yoon20,Baars20}. 
According to our previous work \citep{Hwang18,Hwang19,Yoon20}, it is necessary to identify self-similar structures in order to reveal the asymptotic behaviours of wall turbulence predicted by the AEH because of the multi-scale nature of turbulence.
As shown in \S \ref{sec:2D}, the large scales ($\lambda_x > 12y$) contained in WASS exhibit self-similarity, and in particular the spectral ridges follow $\lambda_x = 4\lambda_z$ over the range (\ref{eq:eq4}).
Hence, the one-dimensional streamwise and spanwise spectra of WASS are computed over the range of (\ref{eq:eq4}), and the corresponding premultiplied spectra are defined as
\begin{subeqnarray}
\Phi^{1D}(k_x,y) & = & \int_{3y}^{0.8-1\delta}\Phi^{2D}(k_x,k_z,y)\frac{d\lambda_z}{\lambda_z},\\[3pt]
\Phi^{1D}(k_z,y) & = & \int_{12y}^{3-4\delta}\Phi^{2D}(k_x,k_z,y)\frac{d\lambda_x}{\lambda_x}.
  \label{eq:eq5}
\end{subeqnarray}
Here, the upper ends of the TBL and the internal flows are $\lambda_x = 3\delta$ and $\lambda_x = 4\delta$, respectively, as discussed in \S \ref{sec:2D}.
The corresponding upper ends of the spanwise wavelength are $\lambda_z = 0.8\delta$ and $\lambda_x = \delta$ according to the linear relationship $\lambda_x = 4\lambda_z$ of the spectral ridge.

\begin{figure}
\centerline{\includegraphics[trim=0cm 0.2cm 0cm 0.2cm,width=14cm]{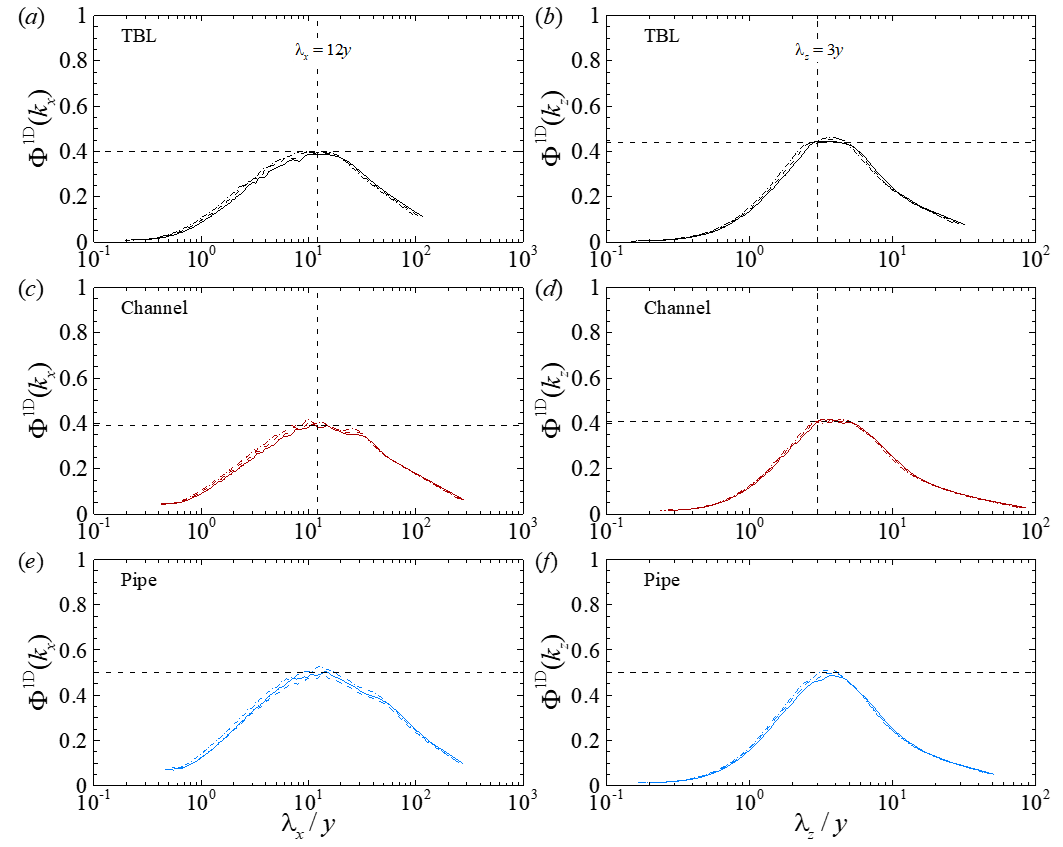}}
\caption{\label{fig:7} Premultiplied one-dimensional streamwise ($a$,$c$,$e$) and spanwise ($b$,$d$,$f$) spectra $\Phi^{1D}$ obtained over the range (\ref{eq:eq4}) at $y^+ = 100$ (solid), $110$ (dashed) and 120 (dashed-dot): ($a$,$b$) TBL; ($c$,$d$) channel; and ($e$,$f$) pipe. Here, each spectrum is normalised by the energy of the fluctuations at a given $y$, obtained by integrating $\Phi^{1D}$. The horizontal dashed lines represent the plateaus or peaks of the respective spectra. The vertical lines denote $\lambda_x = 12y$ ($a$,$c$) and $\lambda_z = 3y$ ($b$,$d$).}
\end{figure}

Figure \ref{fig:7} displays $\Phi^{1D}(k_x)$ (left column) and $\Phi^{1D}(k_z)$ (right column) normalised by the energy of the fluctuations at a given $y$ obtained by integrating each $\Phi^{1D}$.
In the TBL and channel data, we can see a plateau region in both $\Phi^{1D}(k_x)$ and $\Phi^{1D}(k_z)$, implying the possible $k_x^{-1}$ and $k_z^{-1}$ scaling.
The lower limits of these regions are located at $\lambda_x = 12y$ and $\lambda_z = 3y$ (denoted by vertical dashed lines), consistent with those of the linear behaviour in the energetic ridges in (\ref{eq:eq4}).
The plateau appears from $\lambda_x = 12y$ to $\lambda_x = 20y$ in figure \ref{fig:7}($a,c$).
Given $y^+ \approx 100$ and $Re_\tau \approx 1000$, the upper limit of the $k_x^{-1}$ region in the outer unit is $\lambda_x \approx 2\delta$.
Similarly, the $k_z^{-1}$ region is bounded by $3y < \lambda_z < 0.5\delta$ in figure \ref{fig:7}($b,d$).
Interestingly, the range of the $k_x^{-1}$ spectra ($12y < \lambda_x < 2\delta$) is consistent with the bound suggested by \cite{Hwangy15} who examined an artificial channel flow that only resolves turbulent structures at a given spanwise length scale.
This result supports the conjecture that the identification of self-similar motions is required in order to observe the $k_x^{-1}$ scaling region.
Furthermore, such a scaling can be observed even for a low $Re_\tau$ if one extracts energy-containing motions properly.
This claim also reflects the hierarchical nature of self-similar motions in the context of the AEH \citep{Perry82,Hwang18}.
Notably, \cite{Baars20} also investigated the range of the $k_x^{-1}$ region based on the spectral filter obtained from spectral coherent analysis.
Their filter depends on the Reynolds number leading to the dependence of the $k_x^{-1}$ region with $Re_\tau$ and they predicted that such a scaling region can appear for $Re_\tau > 60000$ at $y^+ = 100$.
\jya{\cite{Srinath18} also reported the appearance of the $k_x^{-1}$ region at $y^+ = 100$.
These results give some evidence to support that $Re_\tau^{1/2}$ scaling may not be appropriate for representing the lower bound of the attached eddies \citep{Baars20}.}

According to \cite{Chandran17}, the ratio of the peaks (or plateaus) in the one-dimensional streamwise and spanwise spectra corresponds to the power of the functions that characterise the lower and upper bounds in the two-dimensional spectra.
In other words, the ratio ($m$) is unity when the energy distribution in the two-dimensional spectra is bounded by the linear relationship, and thus the same magnitudes of the plateaus in the one-dimensional spectra indicate the existence of self-similar motions \citep{Chandran17,Deshpande20}.
In figure \ref{fig:7}, the plateau magnitudes of $\Phi^{1D}(k_x)$ and $\Phi^{1D}(k_z)$ are approximately $0.4$ in the TBL and in the channel flow, which also supports the self-similarity in the large-scale region.
In contrast to the TBL and channel data, there is no clear plateau in the pipe data (figure \ref{fig:7}$e,f$).
However, the magnitude of the peak is approximately $0.5$ in $\Phi^{1D}(k_x)$ and $\Phi^{1D}(k_z)$, indicating that the turbulent motions over the range (\ref{eq:eq4}) are responsible for the self-similar behaviour in the pipe flow.
The absence of the $k^{-1}$ scaling in pipe flows may be a consequence of a difference in the flow geometry in which there is less space for wall-attached structures in the pipe flows with increasing $y$ when compared with the TBL or channel flows \citep{Chung15}.
Given $m \approx 1.0$ in all three flows (figure \ref{fig:7}), the present work solely extracts large-scale self-similar motions (\ref{eq:eq4}) contained in WASS.

\jya{The one-dimensional spectrum $\Phi^{1D}$ shown in the present study was obtained by integrating $\Phi^{2D}$ over (\ref{eq:eq4}), where some evidence for self-similar scaling ($\lambda_x \sim \lambda_z$) is observed.
Although the range of the $k^{-1}$ scaling region is relatively narrow owing to the low $Re_\tau$ of our data, the present study shows that such a region appears in a similar subrange of the self-similar scaling in $Ф^{2D}$.
Moreover, the ratio between the magnitudes of the plateau in the one-dimensional streamwise and spanwise spectra is close to unity, supporting that $\Phi^{2D}$ over \ref{eq:eq4} is characterised by $\lambda_x \sim \lambda_z$.
Even at high $Re_\tau$ ($\approx$ 26000), this ratio was found to be 0.79 by \cite{Chandran17} and 0.85 by \cite{Deshpande20}.}

\jya{Given that $\Phi^{1D}$ was computed over (\ref{eq:eq4}), and not the entire range of wavelengths, the integration of $\Phi^{1D}$ does not correspond to the streamwise turbulence intensity $\langle uu \rangle_{ws}$ in figure \ref{fig:4}($b$).
Hence, the magnitude of the plateau in $\Phi^{1D}$ does not match with the slope in figure \ref{fig:4}($b$).
However, the results support the inference that only the energy contained within (\ref{eq:eq4}) contribute to the $k^{-1}$ scaling and the energy contributions from other motions (i.e. the scales that are out of the range (\ref{eq:eq4})) can contaminate the presence of the $k^{-1}$ region.
This contamination could ultimately lead to the absence of the $k^{-1}$ region even when the turbulence intensity appears to follow the logarithmic variation in experiments with a high Reynolds number.
This conclusion aligns with the work of \cite{Davidson06} who argued the aliasing problem in the one-dimensional spectra was due to the shifting of energy to a longer wavelength \cite{Tennekes72}.
In other words, one-dimensional spectra can be contaminated when the spectra are measured over all the wavelengths; e.g. the energy contribution from short $\lambda_z$ ($\lambda_z < 3y$) is included at relatively long $\lambda_x$ (\ref{eq:eq4}$a$) in the one-dimensional streamwise spectra.}

\section{Conclusions}\label{sec:Concl}
We have demonstrated that the wall-attached self-similar structures (WASS) of streamwise velocity fluctuations ($u$) exhibit self-similar behaviour in the context of Townsend's attached-eddy hypothesis (AEH), with special focus on the spectral contribution of turbulence motions contained within the identified structures.
We extract the wall-attached structures of $u$ in the DNS data of a zero-pressure-gradient turbulent boundary layer, and turbulent channel and pipe flows at $Re_\tau \approx 1000$ by identifying the clusters of intense fluctuating regions in the instantaneous flow fields.
The wall-attached structures of $u$ are decomposed into buffer-layer, self-similar, and non-self-similar structures in terms of their height ($l_y$); particular attention is paid to the turbulent statistics contained within the self-similar structures.
The variations in the physical sizes of WASS not only show a good agreement in all three flows but also scale with $l_y$.
In addition, the streamwise turbulence intensity carried by WASS exhibits a logarithmic variation with a similar slope across the logarithmic region.
On the other hand, the sizes of the tall wall-attached structures ($l_y > 0.6\delta$) are characterised by $\delta$ (i.e. non-self-similar) and the corresponding turbulence intensity shows a discrepancy among all three flows with an absence of the logarithmic variation.
We also examine the two-dimensional spectra of $u$ within WASS to explore the spectral signatures of self-similarity proposed by Perry and coworkers.
Across the logarithmic region, the lower and upper bounds of the two-dimensional spectra follow a linear relationship $\lambda_x \sim \lambda_z$ in the large-scale range ($\lambda_x > 12y$).
Moreover, the spectral ridges exhibit $\lambda_x \approx 4\lambda_z$ over the range $12y < \lambda_x < 3-4\delta$, indicating that only the large-scale motions contained in WASS are self-similar.
Based on this spectral band, the one-dimensional streamwise and spanwise spectra are obtained by integrating the two-dimensional spectra.
\jya{They show some evidence for the existence of the $k_x^{-1}$ and $k_z^{-1}$ scaling in a similar subrange identified in the two-dimensional spectra. 
Although the range of such scalings is narrow due to low $Re_\tau$, the magnitudes of the plateau (or peak) in the streamwise and spanwise spectra are close to each other, representing $\lambda_x \sim \lambda_z$.}
Our results support that the asymptotic behaviours of turbulent statistics, predicted by the AEH, can be captured when we adequately filter out contributions from coexisting non-self-similar motions. 
\jya{It is shown that the logarithmic variation in the streamwise turbulence intensity, the self-similar scaling in the two-dimensional spectra, and the possible $k^{-1}$ scaling can be observed simultaneously in the case of instantaneous flow structures extracted by applying the same filter in canonical wall turbulence.}
Given the hierarchical distributions of attached eddies, the self-similar motions can exist even at low $Re_\tau$ although they are not statistically dominant owing to insufficient space in the logarithmic region.
In this respect, the identified WASS are representative energy-containing motions satisfying the AEH and serve as a structural basis for explaining the asymptotic behaviours of wall turbulence. 
However, further study is required to conclude a definite bound \jya{and wall-normal location} of the spectral overlap argument and to associate constants \jya{(e.g. Townsend--Perry constant)} by exploring a wider range of $Re_\tau$.
\\

\noindent \textbf{Acknowledgements}

\noindent This work was supported by the National Research Foundation of Korea (NRF) grant funded by the Korea government (MSIT) (No. 2020R1F1A104853711).\\ 

\appendix
\section{Effect of the structure-identification threshold}\label{appA}
\begin{figure}
\centerline{\includegraphics[trim=0cm 0.2cm 0cm 0.2cm,width=14cm]{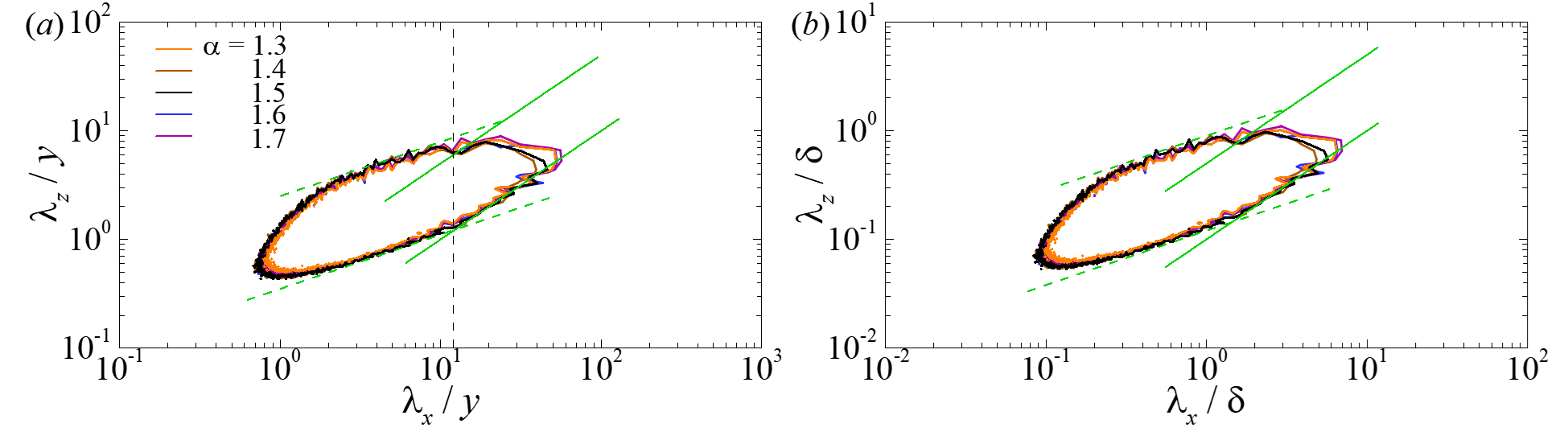}}
\caption{\label{fig:A1} Premultiplied two-dimensional energy spectra $\Phi^{2D}$ at $y^+ = 120$ in TBL with $\alpha = 1.3, 1.4, 1.5, 1.6$, and $1.7$. The green dashed and solid lines are consistent with those in figure \ref{fig:5}. The vertical dashed line indicates $\lambda_x = 12y$.}
\end{figure}
\jya{The wall-attached $u$ structures defined in the present study (\ref{eq:eq1}) depend on the threshold value $\alpha$ because the structures are identified by extracting the physically connected volumes of intense fluctuations.
However, it does not mean that we can use any arbitrary value of $\alpha$ to examine coherent structures.
As $\alpha$ decreases, new structures arise, or some of the previously identifies objects become connected.
The maximum number of structures occurs in the vicinity of $\alpha = 1.5$, indicating that the former behaviour is dominant near $\alpha = 1.5$; see figure 2 in \cite{Hwang18}.
In addition, the maximum volume of the identified object changes significantly near $\alpha \approx 1.5$, which represents the occurrence of the percolation crisis \citep{Moisy04,Del06b,Lozano12,Hwang18}.
In other words, $\alpha \approx 1.5$ effectively captures the intense $u$ structures.
Hence, we examined the influence of the threshold value over a certain range in the vicinity of $\alpha \approx 1.5$.
The threshold effect on the population density and the sizes of the bounding box was reported in \cite{Hwang18}.
In this section, the influence of the threshold on the premultiplied two-dimensional spectra ($\Phi^{2D}$) is shown in figure \ref{fig:A1}.
The value of $\alpha$ varies from 1.4 to 1.7 in the vicinity of the region where the percolation transition occurs \citep{Hwang18}.
Notably, the maximum volume of the wall-attached structures decreases two times from $\alpha = 1.4$ to $\alpha = 1.7$ and the total number of the identified structures varies 10\% over this range.
To avoid any repetition, we plot $\Phi^{2D}$ for TBL only.
For comparison, we plot the green solid and dashed lines consistent with those in figure \ref{fig:5}.
We can see that all the contours collapse reasonably well.
In particular, the linear relationship ($\lambda_x ~ \lambda_z$) appears at $\lambda_x > 12y$, and the lower bounds align along $\lambda_x = 10\lambda_z$ regardless of the threshold.
This supports the inference that the conditionally sampled flow field $u_{ws}$ can represent a continuous range of scales related to the energy-containing motions in the logarithmic region.
Hence, the results remain qualitatively unchanged over the percolation transition region.}




\end{document}